\documentclass[conference]{IEEEtran}
\IEEEoverridecommandlockouts
\usepackage{cite}
\usepackage{tikz}
\usepackage{amsmath,amssymb,amsfonts}
\usepackage{algorithmic}
\usepackage{graphicx}
\usepackage{textcomp}
\usepackage{xcolor}
\usepackage{url, hyperref}
\usepackage{CJKutf8}
\def\BibTeX{{\rm B\kern-.05em{\sc i\kern-.025em b}\kern-.08em
    T\kern-.1667em\lower.7ex\hbox{E}\kern-.125emX}}

\newcommand\copyrighttext{%
  \footnotesize \textcopyright 2024 IEEE. Personal use of this material is permitted.
  Permission from IEEE must be obtained for all other uses, in any current or future
  media, including reprinting/republishing this material for advertising or promotional
  purposes, creating new collective works, for resale or redistribution to servers or
  lists, or reuse of any copyrighted component of this work in other works.
 }
\newcommand\copyrightnotice{%
\begin{tikzpicture}[remember picture,overlay]
\node[anchor=south,yshift=10pt] at (current page.south) {\fbox{\parbox{\dimexpr\textwidth-\fboxsep-\fboxrule\relax}{\copyrighttext}}};
\end{tikzpicture}%
}

\begin{document}
\title{A quest through interconnected datasets:\\
lessons from highly-cited ICASSP papers}

\author{\IEEEauthorblockN{Cynthia C.~S. Liem
\IEEEauthorblockA{\textit{Delft University of Technology}\\
Delft, The Netherlands\\
c.c.s.liem@tudelft.nl\\
https://orcid.org/0000-0002-5385-7695}
\and
\IEEEauthorblockN{Do\u{g}a Ta\c{s}c{\i}lar}
\IEEEauthorblockA{\textit{Delft University of Technology}\\
Delft, The Netherlands\\
dtascilar.cse@gmail.com}
\and
\IEEEauthorblockN{Andrew M. Demetriou}
\IEEEauthorblockA{\textit{Delft University of Technology}\\
Delft, The Netherlands\\
a.m.demetriou@tudelft.nl\\
https://orcid.org/0000-0002-0724-2278}
}}

\maketitle

\copyrightnotice

\begin{abstract}
As audio machine learning outcomes are deployed in societally impactful applications, it is important to have a sense of the quality and origins of the data used. Noticing that being explicit about this sense is not trivially rewarded in academic publishing in applied machine learning domains, and neither is included in typical applied machine learning curricula, we present a study into dataset usage connected to the top-5 cited papers at the International Conference on Acoustics, Speech, and Signal Processing (ICASSP). In this, we conduct thorough depth-first analyses towards origins of used datasets, often leading to searches that had to go beyond what was reported in official papers, and ending into unclear or entangled origins. Especially in the current pull towards larger, and possibly generative AI models, awareness of the need for accountability on data provenance is increasing. With this, we call on the community to not only focus on engineering larger models, but create more room and reward for explicitizing the foundations on which such models should be built.
\end{abstract}


\section{Introduction}
Today, many people have integrated audio services in their daily lives: they listen to music services, and make use of voice assistants. Many of these are enabled by audio machine learning. As with any machine learning task, the quality of a model is highly dependent on that of the dataset provided \cite{quality}. Ideally, for high-stakes applications, data should be responsibly sourced, well-documented, and available for independent quality auditing. However, current publication culture has disincentivized these aspects. In practice, paper authors are not eager to share datasets upon request (even when having promised so in their papers)~\cite{gabelica}, annotation errors exist and impact classification outcomes~\cite{wrong}, and annotation practice standards are missing, causing `garbage in, garbage out' effects in outcomes~\cite{geiger}.

To add to this, doing predictive-focused modeling work within a machine learning pipeline is professionally more valued than working on data quality. This has both emerged in interviews with practitioners~\cite{data_cascades}, as well as from studies into self-reported contributions in highly-cited works at major technical machine learning venues~\cite{values_in_ML}, which mostly highlight performance, generalization, quantitative evidence, efficiency, building on past work, and novelty~\cite{values_in_ML}. As a consequence, this is what peer reviewers typically look for when assessing the merit of a submission, while work seeking to ask more critical questions on the foundations on which work is built has a harder time getting through. Thanks to the OpenReview platform, it can e.g.\ be seen that an earlier submission of~\cite{values_in_ML} was rejected at NeurIPS for being ``not surprising or deep''\footnote{\url{https://openreview.net/forum?id=oioB7Te7Bo}, accessed July 21, 2024.}, while a recent analysis of the ImageNet dataset~\cite{nine_lives_imagenet} also received reviewer criticisms that no new idea or methodology was proposed\footnote{\url{https://openreview.net/forum?id=jh0ck1bPGF}, accessed July 21, 2024.}. More broadly in computer science, concerns have been voiced on whether peer review currently too strongly focuses on novelty, at the cost of gaining bigger-picture insights~\cite{lee2022}.

The work presented in this article follows from these notions, with the first and last author of this work being members of an interdisciplinary lab. Having a background in psychology, the last author has been trained in data acquisition and annotation methods that ensure that \emph{constructs}, i.e.\ phenomena that rely on human interpretation (and cannot directly physically be measured), are measured in valid and reliable ways. For this, the basics of psychometrics~\cite{Furr2014PsychometricsIntroduction} and survey science \cite{groves2009survey} are often taught as entry-level courses to social science students, which inform robust sampling and survey design for gathering human responses. Additionally, due to perverse publication and academic rewarding culture, the psychological research field went through a crisis in which many published claims turned out both irreproducible and irreplicable. Ever since, it has been reforming and transforming to improve credibility~\cite{vazire_credibility}. Nowadays, more rigorous reporting standards are required, and early transparency on what will be researched (e.g.\ through pre-registration) is frequently given, or even expected. Furthermore, the field has generally been displaying more awareness that in order to make general claims, purposeful strategy needs to have been established on whether sufficiently representative samples of the human population are approached, and what questions they are asked to answer. Authoritative psychological publication venues now have base expectations of such strategies existing and being followed, for any work to be submitted.

In contrast, this type of awareness typically has not existed in technical applied machine learning communities. Here, the dominant way of working has primarily focused on engineering technical system innovations that can echo `ground truth' human responses---and as such, may be able to automatically scale up the prediction of human-like responses in various applications. However, the ground-truth responses often are conveniently or sparsely sampled, without clear philosophy on whether they can indeed be considered sufficiently robust and representative to be used as a golden standard. For example, in response to the question, ``How many annotators would be needed for NLP corpus ground truth?'', a well-cited book on natural language annotation for machine learning~\cite{PustejovskyStubbsNLannotation} suggests to ``have your corpus annotated by at least two people (more is preferable, but not always practical)''. This is a remarkably low number, which would not meet the bar of rigor expected today in the quantitative social sciences, and without clear substantiation of whether this indeed would be sufficient.

Some awareness on these issues have been arising in recent years in several works targeting computationally oriented communities, mostly from a responsible computing perspective. For example, several articles raised awareness on how metrology and measurement best practices can be helpful in more robustly dealing with human responses to subjective and more normative questions~\cite{welty_metrology, jacobs_measurement_2021}, and best practices are pointed to with regard to purposeful archiving~\cite{JoGebru2020archives}. In several applied machine learning venues, we do notice calls for increased rigor, accountability and ethical reflection, which typically manifests in mandatory inclusion of ethics statements, or checklists~\cite{Rogers2021responsible} accompanying submissions, as e.g.\ can today be seen at NeurIPS and the *CL venues, which partially are inspired by proposals for better documentation through data sheets\cite{data_sheets} and model cards\cite{model_cards}. Still, institutionalized uptake of the expertise remains relatively rare, and the most highly cited papers at prestigious venues still appear to favor technical innovation over more foundational reflection. As a further recent example, warnings have emerged that key claims in currently popular Large Language Model research often are overblown~\cite{rogersluccioni24}, and many issues exist with regard to test data contamination. In fact, contamination happened for the GPT-3 model, where the authors reported that ``a bug in the filtering caused us to ignore some overlaps, and due to the cost of training it was not feasible
to retrain the model''~\cite{GPT3}.

The first author of this work has a background in music information retrieval. Here, data issues were raised in literature too: faults~\cite{sturm14} and a likely lack of ecological validity~\cite{liemmostert20} in datasets caused for automatically trained classifiers to seemingly perform well in their original evaluations, but---when more thorough testing was applied---not actually pick up the concepts they were purported to capture. Now being a senior lab lead and educator at a public institution, she has sought to raise awareness about these issues with students and peers. However, beyond first-principle argumentation, the first and last author of this paper were frequently requested to deliver more tangible evidence that current common ways of dataset handling indeed are problematic.

Surveys on annotation practices in several applied machine learning domains did emerge, yielding empirical evidence that explicit justification of annotation choices are typically lacking in published work~\cite{geiger, geiger_revisited}, and annotation quality indeed affects model performance~\cite{kern2023annotation}. To add to this body of insight, in Summer 2023, a topic proposal was submitted to the 3rd year undergraduate Research Project in the Computer Science and Engineering curriculum at Delft University of Technology. Here, students were invited to choose an impactful applied machine learning domain or publication venue, and conduct a systematic literature review in which annotation practices were documented for corresponding top-cited publications.

The current paper presents the results of an analysis performed by a then-undergraduate student (the second author), focusing on highly-cited audio research papers. In this particular work, where the original intention had been to conduct a systematic literature review, it turned out non-trivial to get sufficiently in-depth information on the data resources used in the papers. As a consequence, the research project turned into a depth-first investigation into dataset origins. We are aware this is not what typical machine learning venues consider publishable, but will show that these outcomes are relevant, contributing tangible insights on what currently is under-reported in typical academic literature. 

\section{Methodology}
For the current paper, we built upon the student analysis as performed in Summer 2023, but re-verified links and references, updated bibliometrics where relevant, and took care to more explicitly position this work in relation to data-centered and responsible research literature.

In terms of the investigation into dataset origins, the first steps of a systematic literature review were conducted. The International Conference on Acoustics, Speech, and Signal Processing (ICASSP) was chosen as a prestigious and English-language venue, the years 2021 and 2022 were chosen to focus on recent work, and sources were sorted by the number of citations to focus on the most impactful work. For each paper, the intention was to systematically collect annotation practice information for every dataset used to train a machine learning model, following the questions as used in Geiger et al.'s survey on annotation practices~\cite{geiger}. However, often, the origins of datasets turned out unclear. When relevant information was absent in a given paper itself, a deeper inquiry was performed through the paper's references, and by searching for any other available documentation, e.g.\ dataset websites, repository wikis, and README files.  Often, the latter was done through multiple queries into terms, technologies and datasets as mentioned in the papers.  In case no paper or online resource was found providing sufficient information on dataset origins and annotation practice, attempts were undertaken to contact the original authors through their provided e-mails and LinkedIn accounts, although these only rarely yielded responses. In some cases, original resources also were not in English, and neither was their documentation (e.g.~\cite{aishellOriginal} in Chinese and \cite{datatang1,datatang2,datatang3,datatang4,csj1,csj2} in Japanese). In such situations, the second author of this work used Google Translate on the provided documentation text. Furthermore, beyond datasets being used in their original, \emph{initial} reported form, we also frequently encountered situations in which earlier-published datasets were \emph{modified} or \emph{combined}. Finally, the use of \emph{pre-trained} models in a given approach also will bring in knowledge inferred from datasets on which this pre-training was performed.

For each resource, the second author documented whether the style of reporting on data is \emph{explicit} (giving clear attention to the dataset, e.g.\ through dedicated sections or mentions in the abstract) or \emph{implicit} (only naming the name of a dataset in passing), and whether any justification was given for this degree of reporting. Then, for initial datasets, structured notes were taken on whether the name of the original dataset matched that of later references to it; who owned the data; in case of speech, what the content is about and who owned the speech; whether any formal instructions were reported on the recording protocol; whether any formal instructions were reported on the annotation protocol; whether any quality verification happened, and what the dataset size was. For modified and combined datasets, we documented type and details of the modification or combination, and continued searching until reaching initial datasets. For pre-trained models, we recorded the name of the dataset that the model was trained on, and then deepened the search until reaching initial datasets.

Because of the considerable work involved into researching these dataset origins, we limited a full analysis to the top-5 cited papers from 2021-2022, according to Scopus bibliometrics in Spring 2023. Rechecking the bibliometrics a year later for the current manuscript, these articles still are among the top-5 highest-cited ICASSP papers from 2021-2022 (Table~\ref{tab:citationmetrics}).

\begin{table}[!htb]
    \centering
    \begin{tabular}{|c|c|c|c|c|}
    \hline
         Paper & 2023 citations & 2023 rank & 2024 citations & 2024 rank \\
         \hline
         \cite{pap1} & 156 & 1 & 385 & 1\\
         \cite{pap2} & 121 & 2 & 276 & 2\\
         \cite{pap3} & 80 & 3 & 157 & 3\\
         \cite{pap4} & 66 & 4 & 132 & 5\\
         \cite{pap5} & 62 & 5 & 154 & 4\\
         \hline
    \end{tabular}
    \caption{Citation bibliometrics for ICASSP papers studied in this paper (reference dates: May 8, 2023 and April 10, 2024)}
    \label{tab:citationmetrics}
\end{table}

\begin{table*}[htbp]
  \scriptsize
  \centering
  \begin{tabular}{|c|p{2.5cm}|p{10cm}|} 
    \hline
    \textbf{Type of Resource} & \textbf{Style of Reporting} & \textbf{Resource} \\
    \hline
    Selected Literature & Explicit & Attention is all you need in speech separation \cite{pap2}\\
    \hline
    Selected Literature & Implicit & \begin{tabular}[t]{@{}l@{}}
                    SA-Net: Shuffle attention for deep convolutional neural networks \cite{pap1}\\
                    Recent developments on ESPNeT toolkit boosted by conformer \cite{pap3}\\
                    FastPitch: Parallel text-to-speech with pitch prediction \cite{pap4}\\
                    End-to-end anti-spoofing with rawnet2 \cite{pap5}
                  \end{tabular} \\
    \hline
    Initial Dataset & Explicit & \begin{tabular}[p]{@{}l@{}}
                    WSJ0 \cite{wsj0},
                    Datatang \cite{datatang1} \cite{datatang2} \cite{datatang3} \cite{datatang4}, 
                    AISHELL-2 \cite{aishell2}, 
                    CSJ \cite{csj1} \cite{csj2},
                    HKUST \\ Mandarin Telephone Speech \cite{hkustSpeech1} \cite{hkustSpeech2}, 
                    HKUST Mandarin Telephone Transcript\\ Data \cite{hkust}, 
                    LDC Fisher Spanish Speech \cite{fishspeech}, 
                    LDC Fisher Spanish - Transcripts \cite{fishtrans}, \\
                    The CALLHOME Spanish Speech \cite{calltrans2}, 
                    The CALLHOME Spanish Transcripts \cite{calltrans1} \cite{calltrans2},\\ 
                    JSUT \cite{jsut}, 
                    VoxForge English \cite{vox1} \cite{vox2}, 
                    AISHELL-ASR0009 \cite{aishellOriginal}, 
                    LibriVox \cite{librivox}, \\
                    SWITCHBOARD \cite{switchboard1}, 
                    TED Talks \cite{ted}, 
                    FreeSound sound library \cite{freesound}
                  \end{tabular} \\
    \hline
    Initial Dataset & Implicit & \begin{tabular}[p]{@{}l@{}}
                    WordNet \cite{wordnet}, 
                    LibriVox data of Linda Johnson \cite{Lindalibrivox}, 
                    MagnaTune Song Dataset \cite{magnatune}
                  \end{tabular} \\
    \hline
    Modified Dataset & Explicit & \begin{tabular}[t]{@{}l@{}}
                    AMT (for checking ImageNet) \cite{amt}, 
                    WSJ0-2/3mix \cite{wsj023}, AMT for evaluating the\\ FastPitch algorithm \cite{amt}, 
                    Switchboard-1 Release 2 \cite{switchboard21} \cite{switchboard22} ,
                    TED-LIUMv2 \cite{ted21} \cite{ted22}
                  \end{tabular} \\
    \hline
    Modified Dataset & Implicit & \begin{tabular}[t]{@{}l@{}}
                    Aurora-4 \cite{aurora},  
                    LibriSpeech ASR corpus \cite{libri}, 
                    WSJCAM0 \cite{wsjcam0}, 
                    MC-WSJ-AV \cite{mcwsjav1} \cite{mcwsjav2}, \\
                    AMT for translation (Fisher and CALLHOME) \cite{amt},
                    LJSpeech \cite{ljspeech}, \\
                    MagnaTagATune dataset \cite{magnatune} \\
                  \end{tabular} \\
    \hline
    Combined Dataset & Explicit & \begin{tabular}[t]{@{}l@{}}
                    ImageNet-1k \cite{imagenet1} \cite{imagenet2}, 
                    Fisher and CALLHOME \cite{fishhome1} \cite{fishhome2} \cite{fishhome3}, 
                    TED-LIUM Release 3 \cite{ted3}
                  \end{tabular} \\
    \hline
    Combined Dataset & Implicit & \begin{tabular}[t]{@{}l@{}}
                    MS COCO \cite{coco}, 
                    AISHELL-1 \cite{aishell11}, 
                    4th CHiME \cite{chime},
                    The REVERB \cite{reverb1} \cite{reverb2},\\
                    ASVspoof2019 \cite{asv1} \cite{asv2} \cite{asv3}
                  \end{tabular} \\
    \hline
    Pre-trained Model & Explicit & \begin{tabular}[t]{@{}l@{}}
                    ResNet50 \cite{resnet}, 
                    Deep Voice 3 \cite{deepvoice}
                  \end{tabular} \\
    \hline
    Pre-trained Model & Implicit & \begin{tabular}[t]{@{}l@{}}
                    Kaldi framwork for WSJ0 \cite{kaldiwsj0}, 
                    Kaldi HKUST recipe \cite{kaldihkust},
                    Tacotron 2 \cite{tacotron} \\
                    WaveNet \cite{wavenet1} \cite{wavenet2}, 
                    Tag A Tune \cite{tagatune} \\
                  \end{tabular} \\
    \hline
    
  \end{tabular}
  \caption{Overview of all investigated resources relating to the papers surveyed in our study.}
  \label{tab:my_table}
\end{table*}

\section{Results and discussion}
All resources found through our search are summarized in Table~\ref{tab:my_table}. Our full research notes are publicly released with this paper~\cite{supplemental_material}.
As can be seen, we notice many implicit styles of reporting. While this improves for initial datasets, several datasets are even in their initial form not explicitly described. Furthermore, while initial datasets frequently report on audio recording protocols, much less explicit reporting is done on whether there were annotator instructions, and the degree of quality verification starkly differs.

To visualize the paths from a main paper down to initial datasets, we furthermore drew graphs to visualize connections between the found resources.
As this led to entangled and interconnected graphs, we now show and discuss main observations for each of the surveyed papers.

\begin{figure}[htb!]
  \centering
  \includegraphics[width=0.5\textwidth]{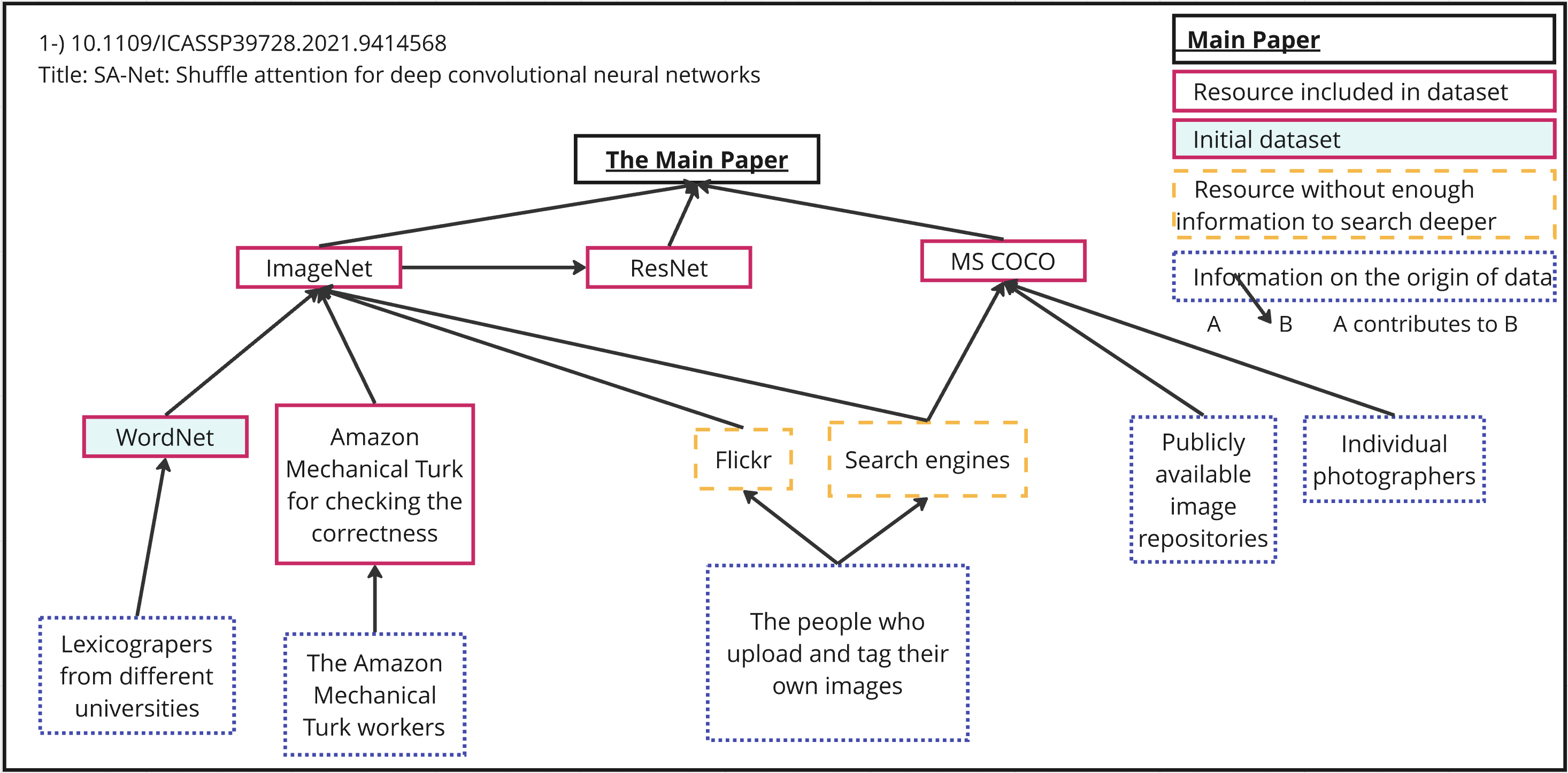}
  \caption{Dataset connections for~\cite{pap1}.}
  \label{fig:pap1}
\end{figure}

The highest-cited ICASSP paper from 2021/2022 is~\cite{pap1}, for which the graph is shown in Figure~\ref{fig:pap1}. The paper proposes a novel Shuffle Attention (SA) mechanism, in which spatial and channel attention mechanisms are combined. Performance is compared to the ResNet model, using ImageNet data. Additionally, experiments on object detection and instance segmentation are reported on MS COCO. Following the paper reporting paths as found through our depth-first search methodology, for ImageNet, several origins of data (Flickr, search engines) are not further traceable. Furthermore, some ambiguity can be found with respect to ResNet: in some experiments, ResNet-50 and ResNet-101 are used as a backbone for the SA mechanism, while it is unclear whether this would be the architecture or a pre-trained model (which commonly would have been done on ImageNet).

\begin{figure}[htb!]
  \centering
  \includegraphics[width=0.5\textwidth]{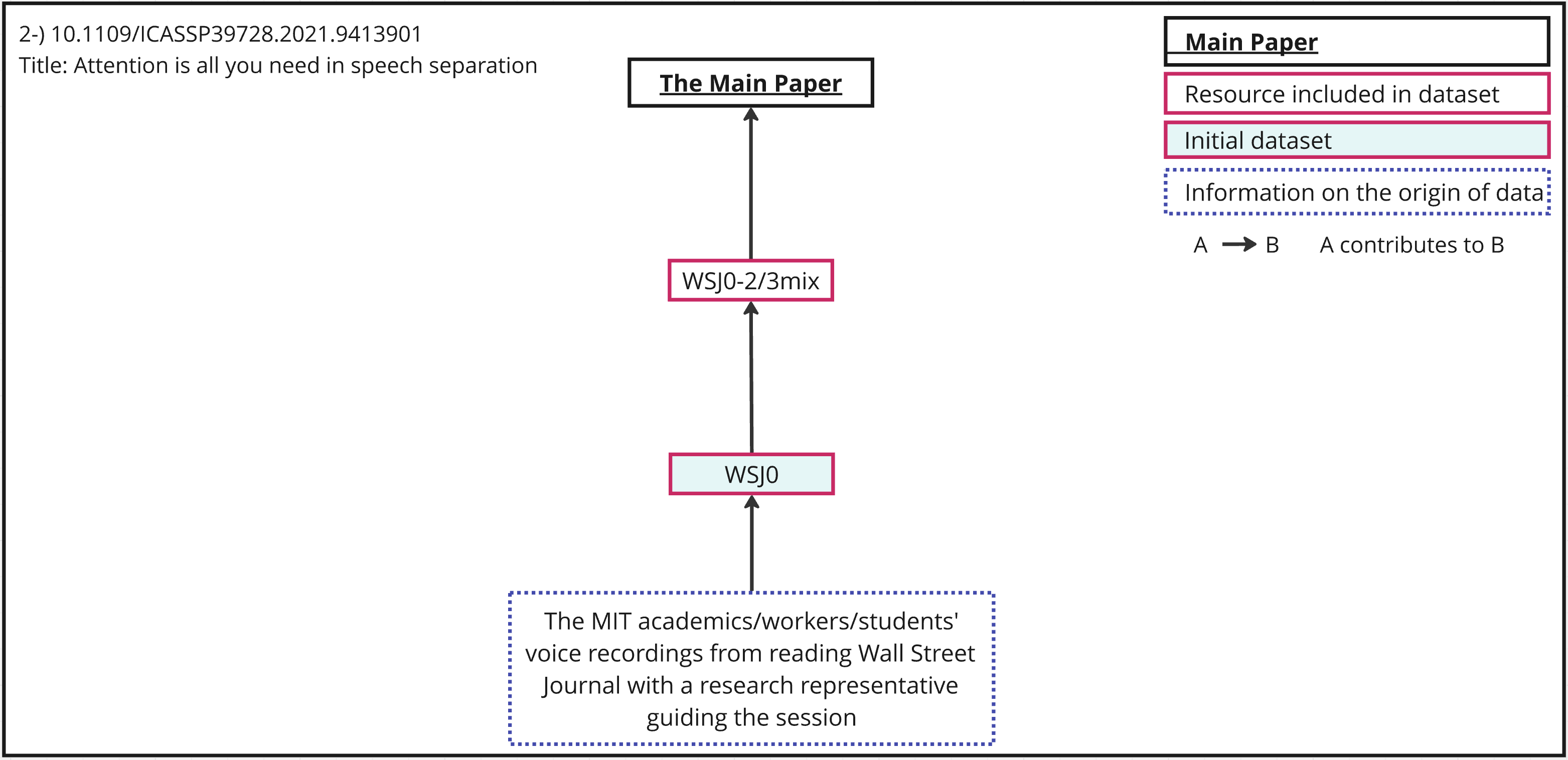}
  \caption{Dataset connections for~\cite{pap2}.}
  \label{fig:pap2}
\end{figure}

Paper~\cite{pap2} proposes a new RNN-free Transformer-based neural network for speech separation, called SepFormer. For experiments, a WSJ0-2/3mix is used, which sources from the WSJ0 dataset. ‘2’ and `3' in WSJ0-2/3mix refer to mixing 2 or 3 different datapoints of WSJ0 to synthesize multiple-speaker recordings.
While this makes the origin graph (Figure~\ref{fig:pap2}) straightforward, validity concerns could be raised on this way of using data. WSJ0 was published in 1991, containing readings from the Wall Street Journal. This will bias vocabulary towards that used in American news of more than 30 years ago. Furthermore, while the way of synthesis is well-controlled, it may not yield realistic multiple-speaker data. Such realistic data actually exists, e.g.\ in the Spanish-spoken Fisher and CALLHOME datasets~\cite{fishhome1, fishhome2, fishhome3}. Finally, we would expect newspaper readings to be more monotonous than the spontaneous speech on which speech separation would more likely be applied.

\begin{figure*}[htb!]
  \centering
  \includegraphics[width=\textwidth]{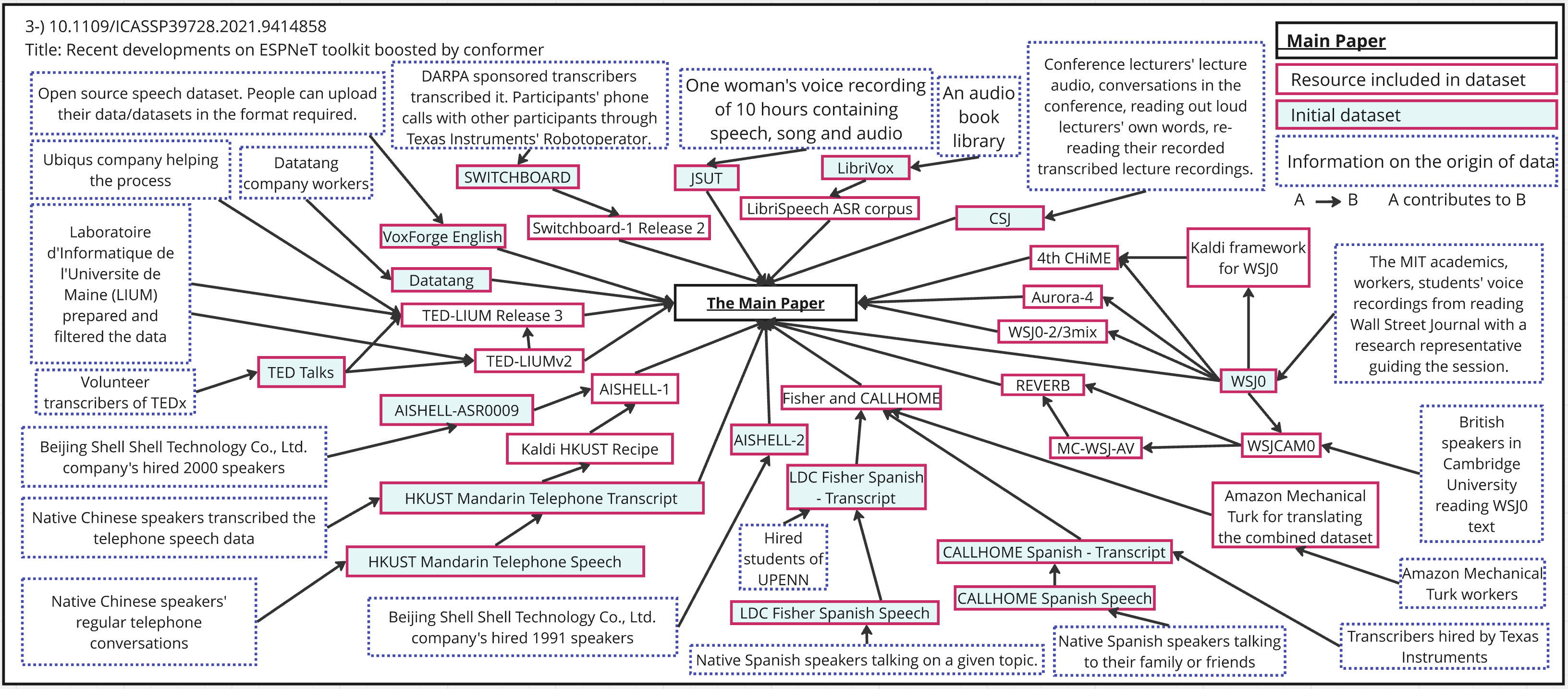}
  \caption{Dataset connections for~\cite{pap3}.}
  \label{fig:pap3}
\end{figure*}

As shown in Figure~\ref{fig:pap3}, a very complicated graph was found for paper~\cite{pap3}, reporting on developments on the ESPnet End-to-end Speech Processing toolkit. This paper reports on the inclusion of a Conformer (convolution-augmented transformer) architecture, and reports on multiple end-
to-end speech processing applications, as should be reflected by 25 automated speech recognition corpora, a speech translation corpus, a speech separation corpus, and three text-to-speech corpora. However, the graph following from our research unveils that these datasets have substantial overlaps. For example, Aurora-4, 4th CHiME and WSJ0-2/3mix all come from the WSJ0 dataset. Likewise, half of TED-LIUM Release 3 is the same as all of the TED-LIUMv2 dataset, and the AISHELL-1 dataset was created by modifying HKUST Mandarin Telephone Transcript Data. Comparing the data resources, they also have considerable disbalances. For example, JSUT represents 10 hours of one woman’s voice recording, while LibriSpeech ASR corpus \cite{libri} represents the sound data gathered from an audio book website having more than 100 speakers with long content. This can lead to undesired biases to be passed on to a model~\cite{quality}. Next to this, it is striking that `multi-lingual' speech datasets  heavily overrepresent English language. No considerations of bias mitigation were discussed. Thus, while the toolkit and the paper are meant to evidence generalizable applicability of the toolkit, questions can be raised on how generalizable the outcomes really are.

\begin{figure}[htb!]
  \centering
  \includegraphics[width=0.5\textwidth]{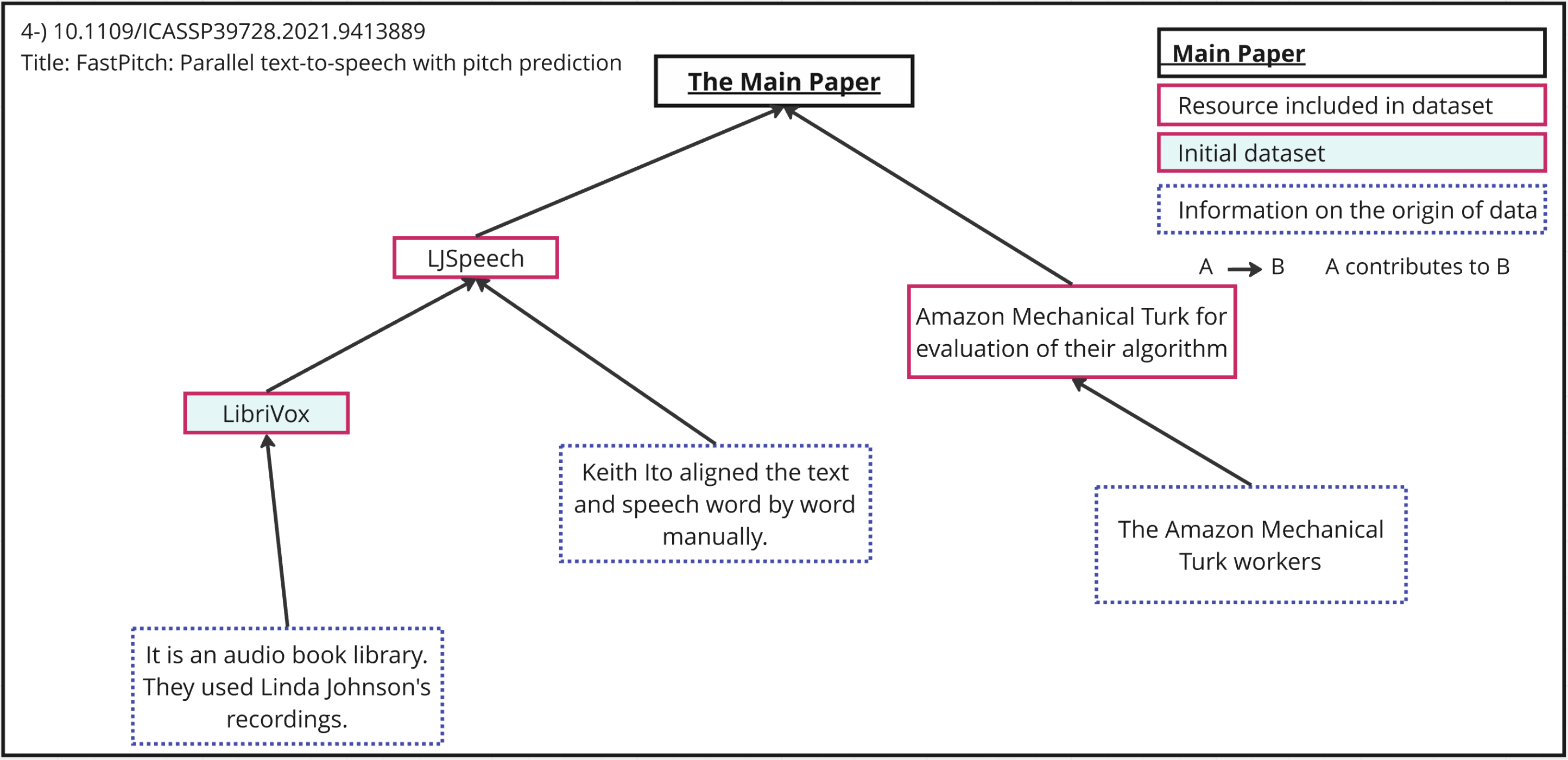}
  \caption{Dataset connections for~\cite{pap4}}
  \label{fig:pap4}
\end{figure}

Paper~\cite{pap4} presents FastPitch, a fully-parallel text-to-speech model based on FastSpeech. The corresponding graph is shown in Figure~\ref{fig:pap4}. The model is trained on the LJSpeech dataset, and subsequently makes use of Amazon Mechanical Turk for algorithm evaluations. The LJSpeech dataset contains e-book voice recordings by Linda Johnson, to which Keith Ito aligned text and the speech data manually. While we initially could not fully obtain information on this dataset from published text, this was the one case in which we had fruitful and active direct interaction after reaching out to the original creator (Keith Ito), who was easily reachable and promptly responded to additional inquiries informing our analysis.

\begin{figure}[htb!]
  \centering
  \includegraphics[width=0.5\textwidth]{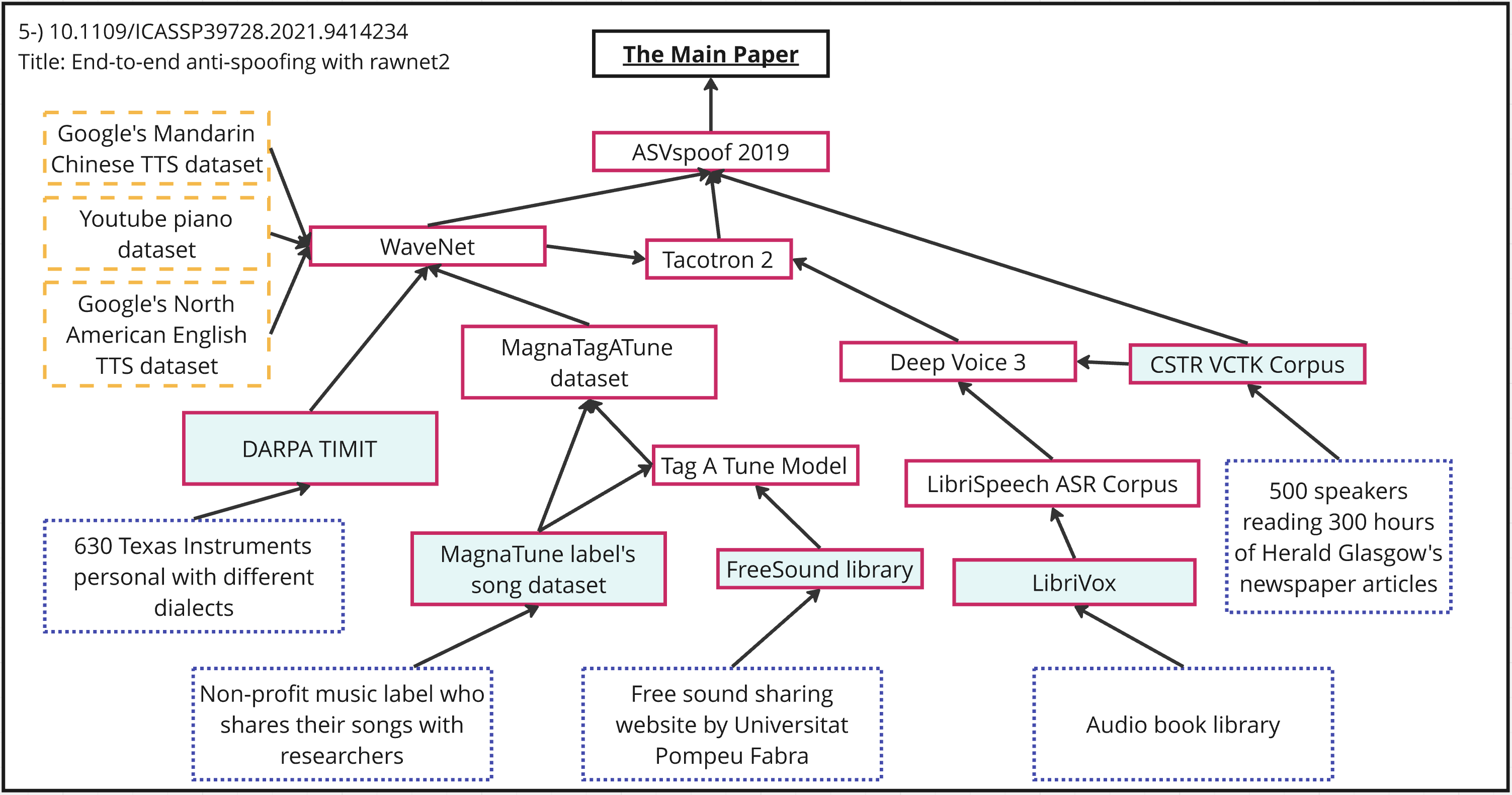}
  \caption{Dataset connections for~\cite{pap5}.}
  \label{fig:pap5}
\end{figure}

Finally, paper~\cite{pap5} describes the application of the RawNet2 deep neural classifier to anti-spoofing in automated speaker verification. The paper mentions using the ASVspoof 2019 logical access (LA) dataset. Looking into this dataset, sparse information is given on its origins. It is derived from the VCTK dataset, and contains bona fide speech and spoofed speech data generated using 17 different text-to-speech and voice conversion systems. The systems used for this generation allegedly also are trained on VCTK data, although it is explicitly mentioned this data does not overlap with data in the ASVspoof 2019 LA dataset. While specification of these 17 systems is reported in~\cite{asv1}, as usual in the domain, the focus most strongly is on describing architectural components, while less attention is paid to explicitizing whether these architectural components may themselves have been trained on data. We noticed explicit mentioning of WaveNet and TacoTron2. WaveNet originally was presented in a more general-purpose way, and as such also incorporates knowledge on (likely copyrighted) music audio. The relations we could trace down are displayed in Figure~\ref{fig:pap5}, but it is likely that a full overview will be more elaborate.
Ultimately, the ASVspoof 2019 LA dataset has been presented as a benchmark dataset and common public resource; as such, it indeed is frequently cited today, although our discussion above once more raises questions on what actually is in the data.

\section{Conclusions and future work}

In this article, we presented a depth-first inquiry into the origins of datasets used in the top-5 cited papers of ICASSP 2021 and 2022. In this, we found noteworthy information on interconnected dependencies, disbalances and unclear origins for the datasets used. This raises questions on the validity and integrity of outcomes trained on these datasets.

Our current work does have several limitations. Because of the substantial amount of deeper searching that was needed to truly get down into the datasets' origins, we only managed surveying five papers. In line with existing surveys~\cite{geiger, geiger_revisited}, in this, we encountered a large degree of implicit reporting of dataset origins and annotation practices. This makes it likely that more links may exist than we reported. Furthermore, the initial search and annotation, most of the verification, as well as the visualizations, were carried out by one researcher (the second author), while independent verification by another researcher (the first author) was only done in case of doubt. As such, the present work may carry annotator bias itself, and our current illustrations of dataset connections should not be read as final and complete overviews, but rather as first information to what data at least may lie underneath a given contribution. 
In releasing our annotations with this manuscript~\cite{supplemental_material}, further interested researchers can independently re-verify them, and possibly expand on them. 

We do wish to acknowledge that in research on audio, speech and music (as well as broader multimedia), data often is copyrighted. As such, only few usable resources may be available, and researchers may for pragmatic reasons be highly incentivized to resort to conveniently available datasets. As we showed in our analysis, this may be problematic. Interestingly, for the papers meant to provide more common shared resources (the end-to-end toolkit in~\cite{pap3} and benchmarking dataset used in~\cite{pap4}), it was more difficult to trace down to dataset origins. Furthermore, while these push towards more generalizable application, still, there always are articulated tasks of primary interest. In such situations, apart from data being larger, having been used for more applications, or being used in state-of-the-art work considered to be trained for general-purpose applications, there is very little explicit justification on whether this aligns with the primarily intended tasks at hand. As has already been raised in literature, it is highly questionable whether datasets aimed at a lot of generic tasks will actually turn out the right choice for application in multiple specialized tasks~\cite{geiger_revisited, whole_wide_world}. Furthermore, it may generally be the case that datasets required for general tasks will require an intractable amount of resources to carefully collect and curate in ways that are representative of general phenomena, rather than specific ones. 

Once again, this historically has not been a focus of interest in our fields. Additionally, many canonical datasets that powered the renewed interest in deep learning (e.g.\ ImageNet, LAION and the Million Song Dataset) are becoming more controversial today~\cite{genealogy_datasets, nine_lives_imagenet, laions_den, kim2023}. Questions arise on whether their content was actually justified and verified, and during the times of dataset and label acquisition, awareness lacked on whether there actually was consent on its use and resharing. Furthermore, resharing practice itself may in situations of copyright only be limited to more privileged parties.

In the current pull towards larger (and possibly generative) machine learning models, again, there is increased demand for more data, while the origins of this data and trustworthiness of corresponding ground-truth annotations may not always be as clear or well-documented. In intended applications, any necessary annotations and outcomes will furthermore to a large degree rely on human judgement.

With this, apart from focusing on more prediction\footnote{or, finding `more' and larger data for broader reuse, while prioritizing size and convenient availability over purposeful quality control}, we strongly call on the applied machine learning research communities to create more room and incentive for purposeful attention to data provenance and annotation quality in academic publishing. We need to set clearer academic integrity examples that normalize paying attention to this in our peer communities. This will need dedicated community effort, that ideally should be backed by senior colleagues, and link to well-established publication venues. Such efforts will naturally lead to more trustworthy bases to depart from---and as such, stronger scientific foundations to the results of our research.

We have seen successful precedents in the establishment of dedicated benchmarking efforts (e.g.\ the MediaEval Benchmarking Initiative for Multimedia Evaluation\footnote{\url{http://multimediaeval.org/}, accessed July 21, 2024.} and the Datasets and Benchmarks track at NeurIPS), and increasing attention to data-centric AI. As examples of concrete possible actions, we can e.g.\ think of focused calls or special sessions to dig more deeply into the origins of commonly used datasets, institutionalized extensions to current checklists and documentation protocols~\cite{Rogers2021responsible, data_sheets, model_cards} with focus on data provenance and ground-truthing procedures, and activities to create better datasets and more robust ground-truth annotations. Alternatively, if applied machine learning communities consider the process of ground-truthing out of scope, it may need to be established as a dedicated academic field.

Finally, we wish to point out that attention to good practices in dealing with human subjectivity in data acquisition and annotation is a core concern in psychology education, but lacking in computing curricula. Many examples in introductory machine learning courses depart from theoretical datasets (e.g.\ synthesized data drawn from known distributions, where parameters are known and controllable), or relate to examples from the natural sciences (e.g.\ iris petal length measurements, where increasing the sample size will indeed be sufficient to get better measurements). However, in reality, many real-world applications will have degrees of ambiguity and reliance on subjective human judgement. As such, we plead for more attention to this in computing education, so future generations of professionals will have sufficient awareness of this.

\section*{CRediT author statement}
Following the CRediT Contributor Roles Taxonomy, we list author contributions in descending order of degree of contribution: \textbf{Conceptualization} AD, DT, CL; \textbf{data curation} DT; \textbf{formal analysis} DT; \textbf{investigation} DT; \textbf{literature review} DT, CL; \textbf{methodology} DT, AD, CL; \textbf{project administration} DT, AD; \textbf{supervision} AD, CL; \textbf{validation} CL, DT; \textbf{visualization} DT; \textbf{writing - original draft} CL, DT; \textbf{writing - review and editing} CL, DT, AD.

\bibliography{references}
\bibliographystyle{IEEEtran}

\end{document}